\documentclass[a4paper,num-refs]{preprint}

\journal{general}

\usepackage{graphicx}
\usepackage{siunitx}

\usepackage{xcolor}
\definecolor{red}{rgb}{1, 0, 0}

\usepackage{pifont}
\newcommand{\cmark}{\ding{51}}%
\newcommand{\xmark}{\ding{55}}%
\definecolor{darkgreen}{RGB}{0,128,0}
\definecolor{darkyellow}{RGB}{204,153,0}
\definecolor{darkred}{RGB}{153,0,0}

\title{BrisT1D Dataset: Young Adults with Type 1 Diabetes in the UK using Smartwatches}

\author[1,\authfn{1}]{Sam Gordon James}
\author[1]{Miranda Elaine Glynis Armstrong}
\author[1]{Aisling Ann O'Kane}
\author[1]{Harry Emerson}
\author[1,\authfn{1}]{Zahraa S. Abdallah}

\affil[1]{University of Bristol}

\authnote{\authfn{1}sam.james@bristol.ac.uk; zahraa.abdallah@bristol.ac.uk}

\papercat{Data Note}

\runningauthor{Gordon James et al.}

\jvolume{}
\jnumber{}
\jyear{}

\begin{document}

\begin{frontmatter}
\maketitle
\begin{abstract}
    \textbf{Background:} Type 1 diabetes (T1D) has seen a rapid evolution in management technology and forms a useful case study for the future management of other chronic conditions. Further development of this management technology requires an exploration of its real-world use and the potential of additional data streams. To facilitate this, we contribute the BrisT1D Dataset to the growing number of public T1D management datasets. The dataset was developed from a longitudinal study of 24 young adults in the UK who used a smartwatch alongside their usual T1D management. \textbf{Findings:} The BrisT1D dataset features both device data from the T1D management systems and smartwatches used by participants, as well as transcripts of monthly interviews and focus groups conducted during the study. The device data is provided in a processed state, for usability and more rapid analysis, and in a raw state, for in-depth exploration of novel insights captured in the study. \textbf{Conclusions:} This dataset has a range of potential applications. The quantitative elements can support blood glucose prediction, hypoglycaemia prediction, and closed-loop algorithm development. The qualitative elements enable the exploration of user experiences and opinions, as well as broader mixed-methods research into the role of smartwatches in T1D management.
\end{abstract}


\end{frontmatter}



\section{Context}


Type 1 Diabetes (T1D) is a medical condition that requires consistent management, which places a significant burden on those who live it. However, the rapid development of technology used to manage the condition is reducing this burden. Continuous Glucose Monitors (CGMs) give increased insight into blood glucose fluctuations \cite{cgm_meta, cgm_trial_ya}, while insulin pumps provide increased flexibility to the user \cite{pump_meta}. Algorithms that link these devices have been developed to automate part of the management process, forming closed-loop artificial pancreas systems. Such technology has been shown to improve both glycaemic outcomes and quality of life for users \cite{closed-loop_meta}, however, further advancements are needed to reach a point of minimal user involvement. Several challenges exist, including the lag that exists both in blood glucose readings and insulin absorption in the body, and the multitude of factors that impact blood glucose levels that the system is unaware of without user input \cite{cl_limits}. Common scenarios that prove challenging to closed-loop systems include meals and physical activity. Exploration into how some of these problems can be addressed requires datasets that include additional data sources, collected in real-world settings that capture the unpredictability of life \cite{t1d_life_transitions}, as done here by the BrisT1D Dataset.

There is a steadily growing group of real-world T1D datasets that are accessible to researchers, with notable examples including OhioT1DM \cite{ohiot1dm}, T1DEXI \cite{t1dexi}, DiaTrend \cite{diatrend}. An overview of these datasets against specific criteria is shown in Table~\ref{tab:datasets}. The OhioT1DM dataset features 12 participants who used a CGM, insulin pump and smartwatch to collect heart rate, accelerometer, galvanic skin response, and temperate data, over 8 weeks. The T1DEXI dataset features 497 participants, 409 of which used a CGM, insulin pump and a study watch to collect heart rate data over 4 weeks. The DiaTrend dataset features 54 participants who used a CGM and insulin pump over a mean of 152 days. In comparison, the BrisT1D Dataset comprises of two elements, `transcripts' and `device data', collected in a six-month longitudinal study from 24 young adults from the UK who used a smartwatch alongside their typical T1D management. The transcripts are from monthly interviews or focus groups with the participants during the study, and the device data is from the CGMs, insulin pumps and smartwatches the participants used. The BrisT1D Dataset complements the other datasets and extends them in several aspects. These points of difference include:
\begin{itemize}
    \item a six-month collection period to enable the assessment of longer term trends.
    \item the inclusion of qualitative data, to allow consideration into user perspectives alongside quantitative analysis.
    \item the inclusion of smartwatch data, including heart rate, steps and distance.
    \item part of the dataset being published open access and available without a request process.
    \item a focus on the young adult population, who has a higher engagement in new technology at a changeable stage in life \cite{ya_tech_use}.
    \item user choice in the smartwatch used in the study to better fit with their wants and requirements, and real-world usage.
    \item variation in CGM and insulin pump device makes and models to highlight further user choice.
\end{itemize}

\begin{table*}[bt!]
    \centering
    \caption{Type 1 Diabetes (T1D) datasets and the criteria they meet. \textcolor{darkgreen}{\cmark}, \textcolor{darkyellow}{$\approx$\cmark}, and \textcolor{darkred}{\xmark} represent passed, partially passed, and failed criteria, respectively. The DiaTrend and T1DEXI datasets have part of the participants using a Continuous Glucose Monitor (CGM) and insulin pump. *The DiaTrend dataset has a vast variation in days of data for each device and across participants, so these figures represent insulin pump data, and the study period is the mean value.}
    \label{tab:datasets}
    \begin{tabularx}{\linewidth}{L R R R C C C}
        \toprule
        \textbf{Dataset Name} & \textbf{Number of Participants} & \textbf{Study Period} & \textbf{Days of Data} & \textbf{CGM and Insulin Pump Use} & \textbf{Activity Data} & \textbf{Openly Accessible} \\
        \midrule
        OhioT1DM & 12 & 8 weeks & 672 & \textcolor{darkgreen}{\cmark} & \textcolor{darkgreen}{\cmark} & \textcolor{darkred}{\xmark} \\
        DiaTrend & 54 & 152 days* & 8,208* & \textcolor{darkyellow}{$\approx$\cmark} & \textcolor{darkred}{\xmark} & \textcolor{darkred}{\xmark} \\
        T1DEXI & 497 & 4 weeks & 13,196 & \textcolor{darkyellow}{$\approx$\cmark} & \textcolor{darkgreen}{\cmark} & \textcolor{darkred}{\xmark} \\
        BrisT1D & 24 & 6 months & 3,903 & \textcolor{darkgreen}{\cmark} & \textcolor{darkgreen}{\cmark} & \textcolor{darkyellow}{$\approx$\cmark} \\
        \bottomrule
    \end{tabularx}
\end{table*}

\section{Methods}

Ethical approval for the study was received from the University of Bristol Engineering Faculty Research Ethics Committee (Ref: 13065). Within the datasets are versions of the participant information sheet and consent form used in the study.

\subsection{Participants}

Participants were recruited through social media posts shared by the T1D charity Breakthrough T1D (formerly JDRF), who posted it on their social media channels, and through emails to participants involved in the authors' previous research who had expressed interest in future research involvement. These recruitment methods were chosen to utilise an existing group of engaged participants and open it to a wider group across the UK to try and maximise engagement in the longitudinal study. 

The inclusion criteria required participants to be 18-26 years old, have T1D, which they had been self-managing for at least a year, using a CGM and insulin pump (with or without closed-loop functionality). The young adult population was chosen as it is an age associated with life transitions and changes that complicate T1D management, which provides a more robust test of the technology \cite{t1d_life_transitions}, and has a higher engagement with new technology compared to other adult age groups \cite{ya_tech_use}. The requirement of CGM and insulin pump usage was put in place to make the dataset more applicable to closed-loop research, a system that has these devices as components.

To register, participants read through an online form that included the participant information sheet, which informed them of what their involvement meant and what the study was aiming to use their data for, and then a consent form to confirm they understood the inclusion criteria and agreed to the study protocol. After the consent form had been signed, participants filled in an online form asking for demographic information through open text boxes. This form was used to understand the diversity of the participant pool and details of their T1D management devices, to check they met the study's requirements. This information is shown in Tables~\ref{tab:demographics}~\&~\ref{tab:technology_use} respectively, with spelling corrected and numbers rounded for consistency where necessary. Those who were found to be suitable participants were invited to an introductory interview, and those who were not suitable were informed and their collected data was deleted. Ten (41.6\%) participants had used a smartwatch before the study.

\begin{table}[bt!]
    \caption{Participants' demographic information.}
    \label{tab:demographics}
    \begin{tabularx}{\linewidth}{L r l l R}
        \toprule
        Participant Number&Age&Gender&Ethnicity&Years since Diagnosis\\
        \midrule
        P01&24&Female&White British&15\\
        P02&24&Female&White British&11\\
        P03&24&Non-Binary&White British&19\\
        P04&21&Female&White British&5\\
        P05&26&Female&White Irish&15\\
        P06&21&Male&White British&19\\
        P07&22&Female&White British&18\\
        P08&20&Female&White British&10\\
        P09&19&Female&White&2\\
        P10&18&Male&White&2\\
        P11&23&Female&White British&5\\
        P12&23&Female&White British&9\\
        P13&21&Female&White Scottish&16\\
        P14&19&Female&British&7\\
        P15&22&Female&Indian&11\\
        P16&21&Female&British&10\\
        P17&26&Female&White&15\\
        P18&22&Female&White British&15\\
        P19&21&Male&White British&20\\
        P20&26&Female&White British&18\\
        P21&24&Female&White&15\\
        P22&20&Female&White British&10\\
        P23&21&Female&White&13\\
        P24&21&Non-Binary&White British&11\\
        \bottomrule
    \end{tabularx}
\end{table}

\begin{table*}[bt!]
    \caption{Participants' Type 1 Diabetes (T1D) technology and smartwatch usage. P06, P07, P11, P18, and P21 upgraded devices during the study, in all cases to a closed-loop enabled system. P17 and P18 were given a new smartwatch but reverted to previously owned devices for the study.}
    \label{tab:technology_use}
    \begin{tabularx}{\linewidth}{L L L L L}
        \toprule
        Participant Number&Continuous Glucose Monitor&Insulin Pump&Closed-Loop Enabled&Smartwatch\\
        \midrule
        P01&Freestyle Libre 2&Medtronic MiniMed 640G&No&Fitbit Luxe\\
        P02&Dexcom G6&Tandem t:slim X2&Yes&Apple Watch Series 5\\
        P03&Dexcom G6&Tandem t:slim X2&Yes&Fitbit Sense\\
        P04&Dexcom G6&Tandem t:slim X2&Yes&Fitbit Versa 4\\
        P05&Freestyle Libre 2&Medtronic MiniMed 640G&No&Fitbit Versa 4\\
        P06&Freestyle Libre 2 \& Dexcom G6&Omnipod Eros \& Omnipod 5&No \& Yes&Fitbit Luxe\\
        P07&Dexcom G6&Omnipod Eros \& Omnipod 5&No \& Yes&Apple Watch Series 5\\
        P08&Dexcom G6&Tandem t:slim X2&Yes&Fitbit Luxe\\
        P09&Dexcom G6&Omnipod Dash&No&Fitbit Sense\\
        P10&Dexcom G6&Tandem t:slim X2&Yes&Fitbit Versa 4\\
        P11&Dexcom G6&Medtronic MiniMed 640G \& Tandem t:slim X2&No \& Yes&Fitbit Sense\\
        P12&Guardian 4&Medtronic MiniMed 780G&Yes&Fitbit Versa 4\\
        P13&Freestyle Libre 2&Omnipod Dash&No&Apple Watch SE\\
        P14&Freestyle Libre 2&Omnipod Dash&No&Fitbit Charge 5\\
        P15&Dexcom One&Medtronic MiniMed 640G&No&Fitbit Sense\\
        P16&Guardian 4&Medtronic MiniMed 780G&Yes&Fitbit Luxe\\
        P17&Dexcom G6&Omnipod 5&Yes&Apple Watch Series 6\\
        P18&Dexcom One \& Guardian 4&Medtronic MiniMed 780G&No \& Yes&Apple Watch Series 5\\
        P19&Dexcom G6&Tandem t:slim X2&Yes&Apple Watch Series 7\\
        P20&Dexcom G6&Tandem t:slim X2&Yes&Fitbit Sense\\
        P21&Freestyle Libre 2 \& Dexcom G6&Omnipod Dash \& Omnipod 5&No \& Yes&Fitbit Luxe\\
        P22&Dexcom G6&Tandem t:slim X2&Yes&Fitbit Versa 4\\
        P23&Dexcom G6&Tandem t:slim X2&Yes&Fitbit Sense\\
        P24&Dexcom G6&Tandem t:slim X2&Yes&Fitbit Luxe\\
        \bottomrule
    \end{tabularx}
\end{table*}

\subsection{Data Collection}

The introductory interview was performed on Microsoft Teams, and during the meetings, the study process was discussed in more detail, and the data extraction processes were tested. Following this, an approximately 30-minute interview was recorded that covered the participants' T1D management, the technology they used, and compared the different smartwatches available to participants as part of the study, which included the Fitbit Sense, Fitbit Versa 4, Fitbit Charge 5, and Fitbit Luxe. The participants then selected one of these options or opted to use a smartwatch they already owned, provided it collected heart rate and step count data, which could be exported from the device. Participants who chose one of the smartwatches offered as part of the study, were sent the device via post and were provided links to set-up tutorials. 

Fitbit smartwatches were selected for their compatibility with a wide range of smartphones, established brand reputation, and ability to display blood glucose data via third-party applications. Offering multiple options enabled participants to select a device that better aligned with their needs, and exploration of this decision. The smartwatches the participants chose are shown in Table~\ref{tab:technology_use}, with 6 participants (25\%) opting to use a pre-owned device, all of which were a model of Apple Watch. Design, functionality, and software varied across the smartwatches used in the study, giving participants varied experiences that better reflected real-world usage. 

Over the next six months, participants were asked to use the smartwatch as felt natural to them. Participants were not given specific guidelines on how often to wear the smartwatch, allowing its usage to more accurately reflect real-life practices. However, in the study sessions, participants were encouraged to explore the features of the smartwatch and consider the role a smartwatch could play in T1D management, with many appropriating it for this purpose. A notable example of this was setting up the smartwatch to show blood glucose readings, which some participants learned about through focus group discussions. Participants were not told to alter their T1D management as a result of wearing the smartwatch for safety reasons.

After approximately a month and for each of the next six months, participants were invited for an interview or focus group, in the order shown in Table~\ref{tab:int&fg}. These interviews and focus groups were all performed by the first author who lives with T1D. The focus group rounds were split into three sessions, with different participants attending each, except for month 5, where there were only two sessions. The monthly interviews typically lasted up to 40 minutes and took place on Microsoft Teams, while the focus groups typically lasted up to 90 minutes and took place on Zoom, to allow participants to hide their names from other participants if they wished. Participants were encouraged to attend the study sessions, although it was not always possible due to scheduling conflicts and as some participants ceased engagement with the study. The participants' attendance for the interviews and focus groups is shown in Table~\ref{tab:int&fg}.

\begin{table*}
    \caption{Interview and focus group rounds during the study period with the number of attendees and the topic(s) covered. The focus group rounds were split into three sessions, with different participants attending each, except for month 5, where there were only two sessions.}
    \label{tab:int&fg}
    \begin{tabularx}{\linewidth}{l r l r l}
        \toprule
        Round&Month&Format&Attendees&Topic(s)\\
        \midrule
        0I&0&Interview&24&T1D management and choosing a smartwatch\\
        1I&1&Interview&21&Initial impression of the smartwatch\\
        2FG[1/2/3]&2&Focus Group&14&Smartwatch data and understanding AI\\
        3FG[1/2/3]&3&Focus Group&14&Reviewing the study so far\\
        4FG[1/2/3]&4&Focus Group&13&Management targets, explainability and data privacy\\
        5FG[1/2]&5&Focus Group&14&Smartwatch metrics\\
        6I&6&Interview&17&Review of the study and smartwatches\\
        \bottomrule
    \end{tabularx}
\end{table*}

Each study round covered a range of different topics around the smartwatch, wider T1D management, and technology, with the topics and example questions from each round shown in Tables~\ref{tab:int&fg}~and~\ref{tab:questions}. The range of topics were chosen to encourage participants to consider the wider implications of smartwatch use in T1D. A combination of interviews and focus groups was employed to gather in-depth individual insights through interviews and foster discussion highlighting smartwatch capabilities during focus groups. 

\begin{table*}
    \caption{Example questions from each of the rounds of interviews and focus groups during the study.}
    \label{tab:questions}
    \begin{tabularx}{\linewidth}{l L}
        \toprule
        Round&Example Questions\\
        \midrule
        0I & What are your opinions on tethered versus patch insulin pumps? How does physical activity impact your T1D management? What smartwatch features do you dislike the idea of? \\
        1I & What smartwatch features have you tried using? How consistently do you wear the smartwatch? How would you improve or adapt the smartwatch to better suit your needs? \\
        2FG[1/2/3] & How has your usage of the smartwatches changed since the first month? How do you think smartwatch data could play a role in T1D management? What would make you more likely to trust artificial intelligence-based T1D technology? \\
        3FG[1/2/3] & What has been the most interesting part for you so far? What smartwatch features are you most commonly using? What changes have you experienced in your daily routine during the study? \\
        4FG[1/2/3] & What would your ideal blood glucose graph look like? How much detail would you want to explain a closed-loop algorithm's actions? Who do you share your T1D data with? \\
        5FG[1/2] & What information from the smartwatch do you look at most often? Which smartwatch metrics would be most useful in predicting future blood glucose levels? Why do you think these metrics would be important? \\
        6I & How did your smartwatch use change over the last 6 months? What advice would you give to someone if they were going to buy a smartwatch? What problems do you see with involving smartwatches in T1D management? \\
        \bottomrule
    \end{tabularx}
\end{table*}

The interviews and focus groups were semi-structured using a topic guide. Several participants commented during the study that the interviewer's personal experience with the condition made them feel more comfortable discussing their own T1D. All the sessions were recorded and transcribed with identifiable information removed, after which the recordings were deleted.

Each month, participants were also asked to export their T1D device and smartwatch data and upload it to a secure OneDrive folder. The platform used to export the device data varied based on the devices participants used but included Glooko, LibreView, Clarity, CareLink, Fitbit Dashboard, Google Takeout and Apple Health. This data included blood glucose readings, basal and bolus insulin doses, carbohydrate intake and physiological data from the smartwatch.

Due to the longitudinal nature of the study \cite{RN745}, six participants disengaged from the study (P08, P09, P13, P14, P20, P23), with only P14 providing reasoning for dropping out of the study, which was due to other commitments requiring their time. The remaining participants who disengaged with the study simply ceased email communication. P08, P14, P20 did so after the introductory interview (0I) and receiving their smartwatch, and P09, P13 and P23 did so after the round 3 focus groups (3FG[1/2/3]), although none attended the preceding rounds of focus groups. Additionally, P15 was unable to find time for any interviews or focus groups after the round 1 interviews (1I), so had no final interview but remained in email contact and submitted device data.

Participants received reimbursement based on their engagement with the study and the cost of their smartwatch. They accrued £$40$ for initial set-up and the introductory interview, £$20$ per monthly interview or focus group they attended, £$10$ for each monthly data upload and £$1$ for each day of data, up to a £$180$ maximum, that had a high data coverage ($\le90\%$ blood glucose data, all insulin data, $\le2$ carbohydrate recordings, and $\le50\%$ smartwatch data). If this total value was less than the value of their smartwatch, participants were given the smartwatch to keep, if the value exceeded the cost of the smartwatch, they received the excess as a bank transfer or voucher in addition to the smartwatch. For example, if a participant chose the FitBit Charge 5, costing £130, and after the initial set-up meeting attended the two interviews and five focus groups, and at each of these meetings uploaded their data, of which 25 of 30 days of each was usable, they would accrue $\text{£}370 (=\text{£}40 + 6\times\text{£}20 + 6\times\text{£}10 + 150\times\text{£}1)$ of reimbursement. This would exceed the initial value of the smartwatch, so they would keep the smartwatch and receive $\text{£}240 (=\text{£}370 - \text{£}130)$ additional reimbursement.

\subsection{Transcripts Processing}

The `transcripts' were first generated using the built-in transcription tools in Microsoft Teams and Word from the recordings of the interviews and focus groups. The first author then corrected these as they listened back over the recordings. During this process and then again before publishing, the first author read through the transcripts and removed identifiable information, including names, locations more specific than countries, building names, and public events the participants had been involved in. Participants were labelled using the randomised participant numbers shown in Table~\ref{tab:demographics}.

\subsection{Device Data Processing}
\label{sec:data_processing}

The `device data', uploaded T1D device and smartwatch data, were processed in two stages. First, the uploaded files were anonymised, with redundant files removed, to form the `raw state'. The raw state was then cleaned, and the frequently occurring metrics aggregated to form the `processed state'. The processed state is designed to be quick and easy to analyse, while the raw state allows for more in-depth exploration of the device data.

\subsubsection{Raw State}

The files uploaded by participants were manually searched for occurrences of potentially identifiable information and empty or redundant files. Where potentially identifiable information was found (for example, names, user notes, device IDs and serial numbers), code was written that deleted the file or removed the column across all exports from the same device platform. Empty or redundant files were also deleted in this process. Additionally, any data from before the start of the study, 1st June 2023, that had been included in the export was also deleted. The exports from Apple Health, the platform used to export data from Apple Watches, included XML files, with higher occurrences of identifiable information, so the relevant activity data was extracted and converted to CSV format for consistency with the other device data. These were split into four files: \texttt{bpm.csv} - additional heart rate data, \texttt{record.csv} - watch sensor metrics, \texttt{tz.csv} - time zones recorded in data, and \texttt{workout.csv} - workout events captured by the watch. The parent file/folder name for each export was changed to reflect the export platform and to feature the date of the most recent data point appearing in it. The exports from each participant were grouped in a unique folder named with their randomly assigned participant number. A CSV file, named \texttt{devices.csv}, containing the devices used by the participants with start dates to highlight the cases where the participant changed devices mid-study. 

A bug was found in the exports from Glooko, which failed to include the duration of extended boluses, which are delivered over longer periods. This impacted 10 participants, and as a result, these durations were manually collected during the final interview and stored in a CSV file, named \texttt{extended.csv}. Additionally, a bug was found in the exports from Glooko from participants who used Omnipod 5 insulin pumps, which meant their basal data was incorrectly recorded. For P7 and P17, who uploaded data while they were using the device, an additional TXT file, named \texttt{basal\_profile.txt}, is included that contains their background basal rates.

\subsubsection{Processed State} 

\begin{figure*}[tb!]
    \centering
    \includegraphics{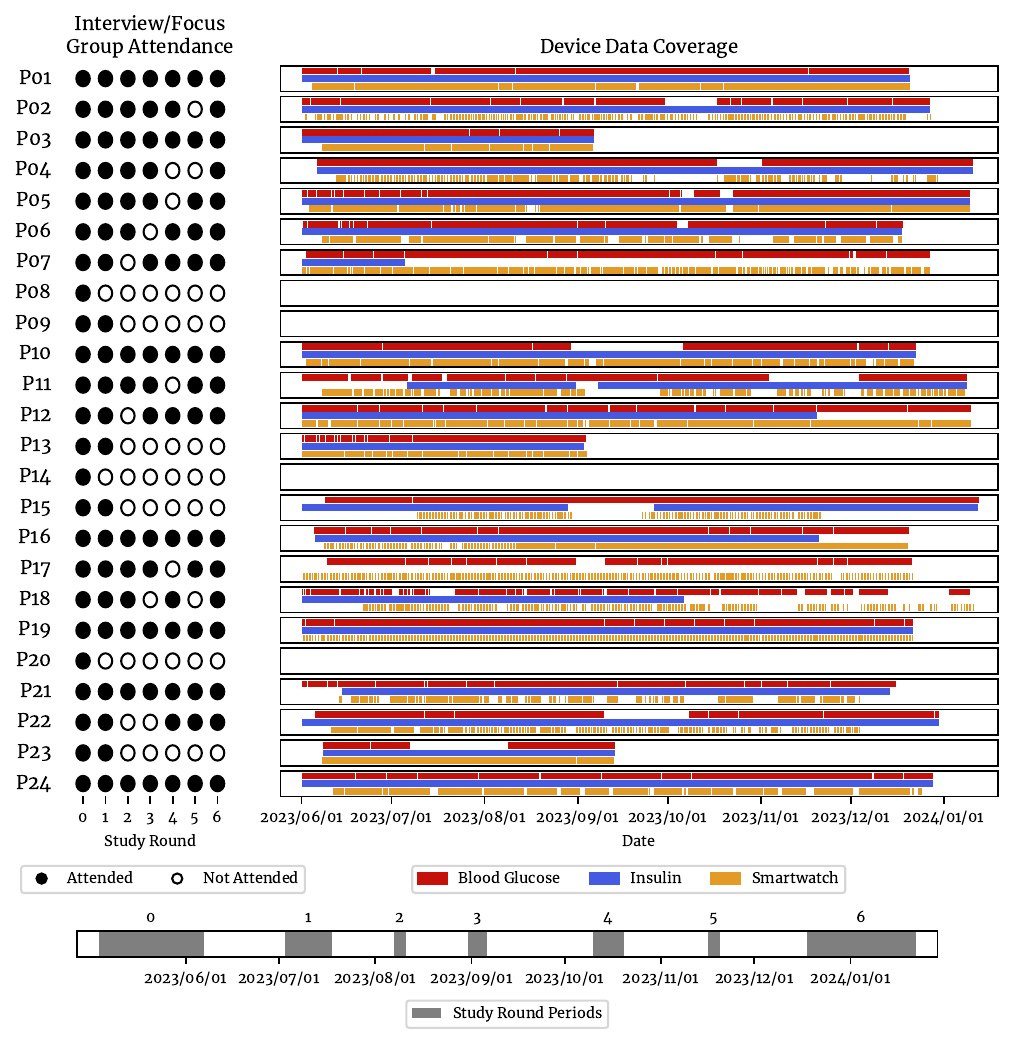}
    \caption{Coverage of the dataset. On the left is the attendance of the participants to the interviews and focus groups. All participants attended the introductory interview in study round 0, in which they chose the smartwatch. Attendance was varied across participants, with 17 attending the final interview in study round 6. On the right is the coverage of the processed state of the device data. Coverage of the data ranges across the participants, with four (P08, P09, P14, and P20) who didn't upload data before they dropped out and other exporting issues limiting the data streams of particular participants (e.g. P17's insulin data).}
    \label{fig:data_coverage}
\end{figure*}

To generate the processed state device data, the following important and consistently occurring metrics were selected and processed:
\begin{itemize}
    \item \textbf{blood glucose level (mmol/L)} - blood glucose readings are rounded to the nearest minute.
    \item \textbf{insulin dosage (U)} - basal rates and extended boluses are assumed to be continuous, and the dose received in the last five minutes is calculated and added to any bolus doses given.
    \item \textbf{carbohydrate intake (g)} - carbohydrate intake records are rounded to the nearest minute
    \item \textbf{heart rate (bpm)} - the mean of any values recorded in the five-minute interval
    \item \textbf{distance (m)} - the total of any values recorded in the five-minute interval
    \item \textbf{steps (count)} - the total of any values recorded in the five-minute interval
    \item \textbf{calories (kcal)} - the total of any values recorded in the five-minute interval
    \item \textbf{activity} - the interval is labelled with the activity if it was being performed for over half of the five-minute interval
\end{itemize}
Other than blood glucose levels and carbohydrate intake, readings were aggregated into five-minute intervals. For this aggregation, the time the data was collected over is divided into 5-minute intervals starting at 00:00 (e.g. 00:00 $\le x_1<$ 00:05 and 00:05 $\le x_2<$ 00:10). This aggregation process and interval length was chosen because machine learning (ML) models struggle with irregular sampling \cite{ml_inconsistent_data}, and to match with the lower bound of the CGM sampling rates. The data from each interval is summed or the mean calculated, depending on the metric. A value is only set for an interval if at least one reading exists within it. The processed data for each participant is combined and saved in a CSV file with the participant number as its name. This process was performed using code written in Python and made publicly available, with more details in the Availability of Source Code and Requirements Section.

Due to the range of platforms the device data was exported from, there is variation in the timezone formatting in the raw state. For Apple Watch data, timestamps were in UTC, so the timezones stored in \texttt{tz.csv} were used to correct the timestamps. For Fitbit data, calorie data was timestamped at local time. However, the heart rate, distance, steps and activity data were timestamped at UTC. Timezone changes were detected using a quirk in the export format that meant while timestamps were UTC they were separated into a different file per day. Therefore, if a participant wore the watch overnight, the difference between the first time that appeared in the file and midnight could be used to calculate the timezone change. Fitbit data timestamped at UTC was then updated using these timestamps before aggregation.

The insulin, carbohydrate, and part of the blood glucose data were recorded on the insulin pumps used, which have no internet connection and therefore rely on the user to maintain the accuracy of the device's clock. This presents the possibility for misalignment in timestamps between data recorded on the insulin pump, compared to any recorded on a smartwatch or phone. For some participants, it was possible to calculate this misalignment when the same blood glucose readings were recorded by the participant's insulin pump and phone. In these instances, the insulin pump data was corrected for this misalignment, but elsewhere it was assumed that the participant was accurately maintaining the insulin pump's clock (which, if it was not the case, may result in some misalignment in readings between devices).

Other processing steps included:
\begin{itemize}
    \item replacing `Low'/`0.1' or `High'/`111.1' values in Dexcom CGM sensor exports with 2.2\,mmol/L and 22.2\,mmol/L, respectively, the upper and lower bounds that, if exceeded, would result in the placeholder values being recorded.
    \item removing duplicate data.
    \item adding the mean extended bolus durations from the manually collected \texttt{extended.csv} data (99 minutes) to any extended boluses missing a duration.
\end{itemize}
Some insulin pumps had erroneous data in part or all of their exports, and so insulin data from these periods were removed. The Omnipod 5 exports did not feature basal rate changes and therefore were removed, which impacted most of P07's and all of P17's insulin data. The format for the exports from participants who used a Medtronic MiniMed 780G with the closed-loop mode enabled changed in the final month of the study, and no longer included the automated changes to insulin delivery made by the insulin pump. As a result, all data from exports from participants using the device after the 1st of December 2023 was removed, which impacted P12's, P16's, and P18's insulin data.

After processing, there were 52,105 hours of data featuring blood glucose, insulin, and smartwatch data. Figure~\ref{fig:data_coverage} depicts the coverage of data from each of the participants in the processed state device data. Four of the participants, P08, P09, P14, and P20, uploaded no or insufficient data to feature in the dataset. Three of the participants, P03, P13, and P23, uploaded around three months of data, and the remainder of the participants uploaded over six months of data. Of these participants, there are gaps within each of the data streams. For the blood glucose and insulin data, these gaps tend to appear in larger chunks, potentially due to misalignment in data uploads or periods of missed data collected. Comparatively, the smartwatch data tends to feature both larger chunks, from consecutive days of not wearing the smartwatch, and small regular gaps, reflecting the daily wear patterns of participants.

\subsection{Dataset Structure}

The BrisT1D is broken into two parts, the BrisT1D-Open Dataset and the BrisT1D-Restricted Dataset, both of which are stored on the University of Bristol Data Repository \cite{databris}. This separation is performed to make a large proportion of the dataset openly available, and therefore easier for researchers to access and use, while restricting unprocessed smartwatch data that carries a higher risk of participant identification \cite{reidentification}. The two parts of the dataset follow the same structural pattern, and at the top level are separated into:
\begin{itemize}
    \item \texttt{device\_data/} - the device data uploaded by participants during the study. (Open and Restricted)
    \item \texttt{study\_forms/} - copies of the participant information sheet and consent form used in the study. (Open and Restricted)
    \item \texttt{transcripts/} - anonymised transcripts of the interviews and focus groups. (Open)
    \item \texttt{demographic\_data.csv} - demographics information of the participants for insights into the diversity of the dataset. (Open)
    \item \texttt{LICENCE.txt} - details of the dataset's licence, Creative Commons Attribution 4.0 \cite{ccby4.0}. (Open and Restricted)
    \item \texttt{README.txt} - details of the dataset. (Open and Restricted)
\end{itemize} 
In the BrisT1D-Open Dataset the \texttt{device\_data} directory contains the \texttt{processed\_state} directory, which in turn contains a CSV for each participant with device data named with their participant number (e.g. \texttt{P01.csv}), with the columns highlighted in Table~\ref{tab:columns}. In the BrisT1D-Restricted Dataset the \texttt{device\_data} directory contains the \texttt{raw\_state} directory, which in turn contains directories named after each participant with device data (e.g. \texttt{P01/}). These participant directories contain all the anonymised device data provided by that participant. For both sides of the BrisT1D Dataset, the \texttt{study\_forms/} directory contains black copies of the participant information sheet and consent form used in the study. Inside the \texttt{transcripts} directory, which is only included in the BrisT1D-Open Dataset, is a directory for each round of the study, containing the transcript from each interview or focus group in that round. 

\begin{table}[tb!]
  \caption{The column names in the \textit{processed} state quantitative data.}
  \label{tab:columns}
  \begin{tabularx}{\linewidth}{l l L}
    \toprule
    Column Name&Unit&Description\\
    \midrule
    timestamp & - & The time and date the reading was taken. For some variables, this corresponds to the end of the interval the data is aggregated over.\\
    bg & mmol/L & Blood glucose level recorded by the continuous glucose monitor.\\
    insulin & U & Total insulin dose received in the previous five minutes from the insulin pump.\\
    carbs & g & Carbohydrate intake recorded by user in the insulin pump or reader.\\
    hr & bpm & Mean heart rate for the previous five minutes as recorded by the smartwatch.\\
    dist & m & Total distance travelled in the previous five minutes as recorded by the smartwatch.\\
    steps & count & Total steps taken in the previous five minutes as recorded by the smartwatch.\\
    cals & kcal & Total calories burned in the previous five minutes as recorded by the smartwatch.\\
    activity & - & Labelled activity events, declared by the user.\\
    device & - & Name of the device that was used to collect the data.\\
    \bottomrule
  \end{tabularx}
\end{table}

\section{Data Validation and Quality Control}


Exploration of the blood glucose, insulin, and smartwatch data took place to assess the validity of the \textit{processed state} device data. Across the figures presented here, a consistent colour scheme is used for the different data streams. Blood glucose data is shown in \textcolor[HTML]{C6110B}{red}, insulin data in \textcolor[HTML]{455AE2}{blue}, carbohydrate data in \textcolor[HTML]{41B041}{green} and smartwatch data in \textcolor[HTML]{E49B25}{orange}. For the blood glucose readings, the 2019 standardised CGM metrics for clinical care were calculated \cite{cgm_metrics, gmi, coef_var}, which are shown in Table~\ref{tab:cgm_metrics}. These metrics highlight a range of blood glucose management with some meeting the clinically set targets and others not, providing a more robust test for those using the data to model real-world management. 

\begin{figure}[tb!]
    \centering
    \includegraphics{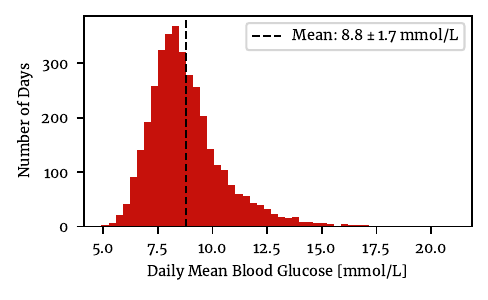}
    \caption{The daily mean blood glucose value across all participants. The skewed normal distribution highlights a large proportion of days have a mean blood glucose 7 and 10 mmol/L, but there are cases of much higher mean glucose levels.}
    \label{fig:mean_bg_hist}
\end{figure}
\begin{figure}[tb!]
    \centering
    \includegraphics{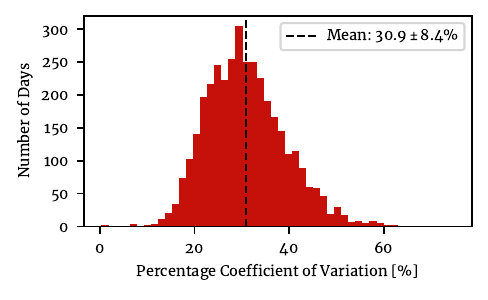}
    \caption{The daily percentage coefficient of variation across all participants, an indicator of glycaemic variability. There is a range of glycaemic variability values over the dataset, approximately normally distributed around 30.9\%.}
    \label{fig:coef_var_bg_hist}
\end{figure}
\begin{figure}[tb!]
    \centering
    \includegraphics{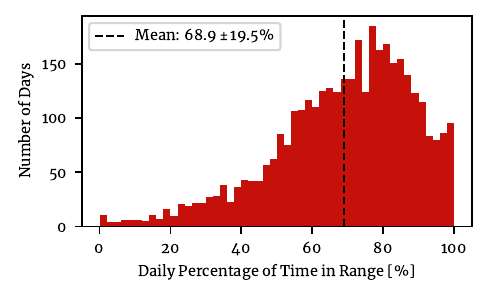}
    \caption{The daily percentage of time spent in the target range (3.9--10.0 mmol/L). The majority of days have a time in range of over 60\%, but there are also cases of much lower time in range.}
    \label{fig:tir_bg_hist}
\end{figure}
\begin{figure*}[tb!]
    \centering
    \includegraphics{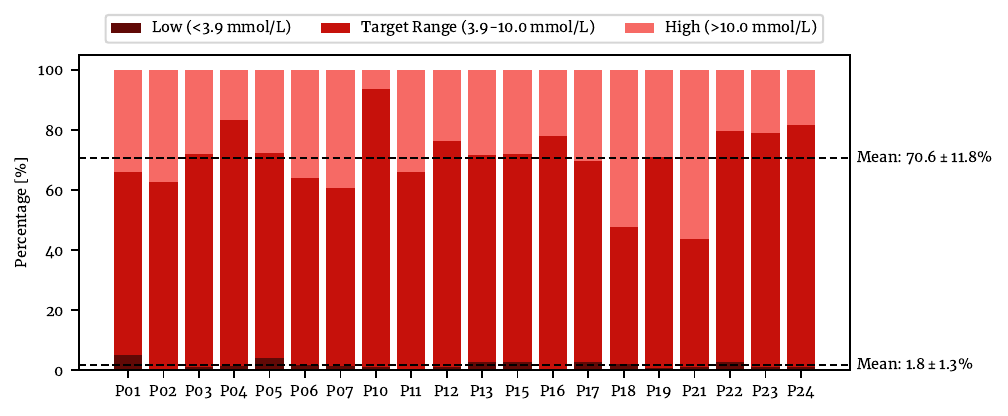}
    \caption{The percentage of time below, in, and above the target range (3.9--10.0 mmol/L) for each participant. The datasets includes examples of participants who have a high time in the target range (P10), and others spend less than 50\% of time in the target range (P18 and P21).}
    \label{fig:tir}
\end{figure*}

The daily values across all participants for mean blood glucose, coefficient of variation and time in range, shown in Figures~\ref{fig:mean_bg_hist}, \ref{fig:coef_var_bg_hist}, and \ref{fig:tir_bg_hist} respectively. These distributions follow expected patterns \cite{diatrend}, with many of the days meeting the target of 70\% time in range but with numerous cases of days falling well below that. The skew of daily mean blood glucose to higher readings is also expected as the more severe short-term impact of low blood glucose tends to lead people to more readily avoid this. The coefficient of variation reflects a near normal distribution around 30.9\% that highlights examples of days with both high and low glycaemic variability. The times above range and times below range have been grouped and presented for each participant in Figure~\ref{fig:tir}, further highlighting the variation across the dataset. When using the dataset, some blood glucose patterns will occur more often in some participants than in others, and so may bias findings to personal effects.

\begin{sidewaystable}
    \caption{Standardised Continuous Glucose Monitor (CGM) metrics for clinical care for each of the participants with device data \cite{cgm_metrics}.}
    \label{tab:cgm_metrics}
    \begin{tabularx}{\linewidth}{l r r r r r r r r r r}
    \toprule
    Participant & Days CGM & Percentage Time & Mean Blood & GMI [\%] & Coefficient of & $x < 3.0$ mmol/L & $3.0 \leq x < 3.9$ & $3.9 \leq x \leq 10.0$ & $10.0 < x \leq 13.9$ & $13.9 < x$ \\
    Number & Worn/Total & CGM is Active [\%] & Glucose [mmol/L] &  & Variation [\%] & [\%] & mmol/L [\%] & mmol/L [\%] & mmol/L [\%] & mmol/L [\%] \\
    \midrule
        P01 & 202/203 & 97.39\% & 8.80~mmol/L & 54.14\% & 44.94\% & 0.71\% & 5.19\% & 60.64\% & 21.55\% & 11.91\% \\
        P02 & 193/210 & 89.57\% & 9.50~mmol/L & 57.42\% & 33.93\% & 0.15\% & 0.42\% & 63.29\% & 25.65\% & 10.49\% \\
        P03 & 98/98   & 98.63\% & 8.56~mmol/L & 52.98\% & 36.35\% & 0.18\% & 1.38\% & 71.33\% & 19.95\% & 7.16\%  \\
        P04 & 205/219 & 92.26\% & 7.81~mmol/L & 49.48\% & 29.49\% & 0.38\% & 1.46\% & 82.48\% & 14.28\% & 1.40\%  \\
        P05 & 217/223 & 92.66\% & 8.37~mmol/L & 52.12\% & 37.91\% & 0.69\% & 4.10\% & 68.05\% & 21.86\% & 5.30\%  \\
        P06 & 198/201 & 94.29\% & 9.40~mmol/L & 56.96\% & 44.29\% & 0.15\% & 1.97\% & 62.25\% & 20.77\% & 14.86\% \\
        P07 & 209/209 & 97.12\% & 9.65~mmol/L & 58.12\% & 39.32\% & 0.42\% & 1.16\% & 60.00\% & 23.60\% & 14.82\% \\
        P10 & 168/205 & 80.51\% & 6.75~mmol/L & 44.47\% & 27.43\% & 0.15\% & 1.37\% & 92.55\% & 5.54\%  & 0.38\%  \\
        P11 & 192/222 & 82.78\% & 8.99~mmol/L & 55.03\% & 30.93\% & 0.05\% & 0.60\% & 66.68\% & 27.74\% & 4.94\%  \\
        P12 & 223/223 & 96.07\% & 8.29~mmol/L & 51.71\% & 37.52\% & 0.19\% & 1.33\% & 75.63\% & 16.99\% & 5.85\%  \\
        P13 & 95/95   & 96.24\% & 8.47~mmol/L & 52.57\% & 34.26\% & 0.21\% & 3.04\% & 69.09\% & 23.36\% & 4.30\%  \\
        P15 & 219/219 & 98.58\% & 8.45~mmol/L & 52.47\% & 36.48\% & 0.57\% & 2.54\% & 69.99\% & 21.36\% & 5.54\%  \\
        P16 & 199/199 & 97.39\% & 8.41~mmol/L & 52.29\% & 22.99\% & 0.01\% & 0.08\% & 79.51\% & 19.99\% & 0.41\%  \\
        P17 & 187/196 & 93.17\% & 8.86~mmol/L & 54.43\% & 34.73\% & 0.71\% & 2.31\% & 67.70\% & 22.59\% & 6.68\%  \\
        P18 & 190/223 & 73.83\% & 10.93~mmol/L & 64.14\% & 41.80\% & 0.71\% & 1.70\% & 46.03\% & 26.81\% & 24.76\% \\
        P19 & 204/204 & 98.82\% & 8.68~mmol/L & 53.54\% & 31.98\% & 0.07\% & 0.71\% & 71.44\% & 23.12\% & 4.66\%  \\
        P21 & 198/198 & 96.56\% & 10.99~mmol/L & 64.44\% & 38.25\% & 0.20\% & 1.32\% & 42.86\% & 32.71\% & 22.91\% \\
        P22 & 181/209 & 85.12\% & 7.88~mmol/L & 49.80\% & 35.08\% & 0.74\% & 2.43\% & 77.20\% & 16.04\% & 3.59\%  \\
        P23 & 67/98   & 65.50\% & 8.03~mmol/L & 50.51\% & 37.91\% & 0.21\% & 1.36\% & 78.20\% & 14.56\% & 5.68\%  \\
        P24 & 211/211 & 97.80\% & 7.89~mmol/L & 49.85\% & 29.94\% & 0.15\% & 1.15\% & 81.11\% & 15.65\% & 1.95\%  \\
    \bottomrule    
    \end{tabularx}
\end{sidewaystable}

\begin{figure*}[tb!]
    \centering
    \includegraphics{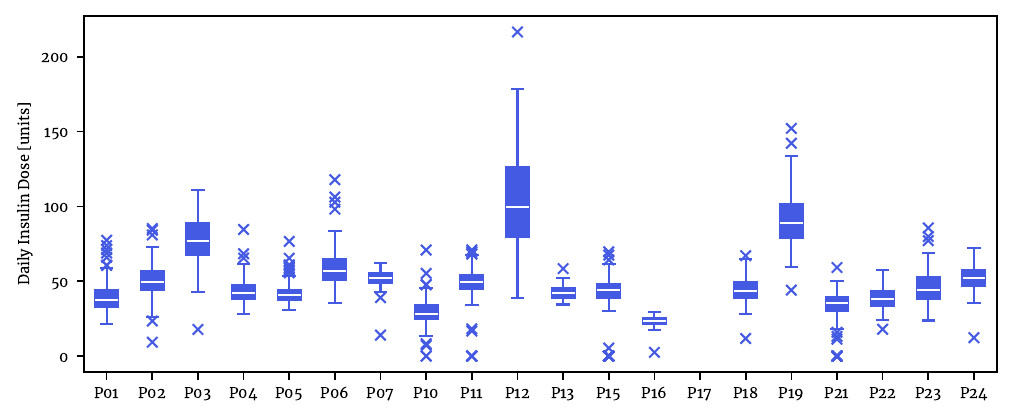}
    \caption{Boxplot of daily insulin dose for each participant. This includes days between the first and last insulin dose featured in the dataset for that participant. P17 has no insulin data due to a problem with the Glooko exports for Omnipod 5 users.}
    \label{fig:insulin_boxplot}
\end{figure*}

\begin{figure*}[tb!]
    \centering
    \includegraphics{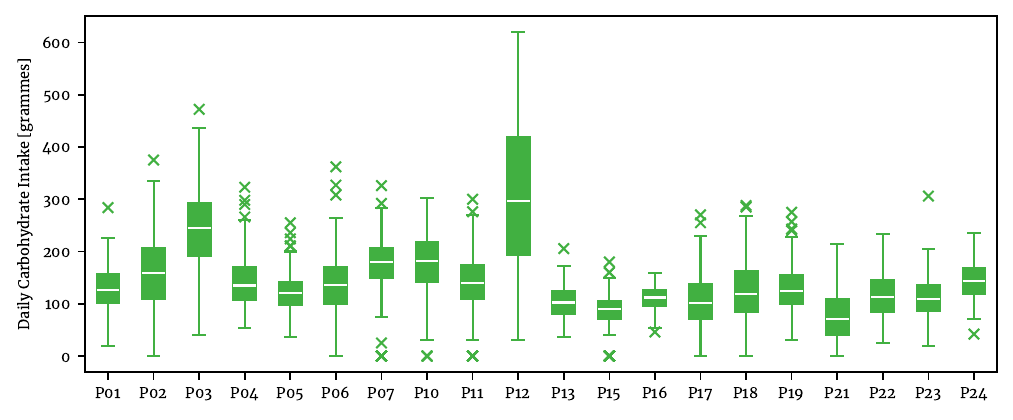}
    \caption{Boxplot of daily carbohydrate intake for each participant. This includes days between the first and last carbohydrate intake featured in the dataset for that participant.}
    \label{fig:carb_boxplot}
\end{figure*}

To explore the insulin and carbohydrate data in the dataset, boxplots were generated for the daily insulin dose and carbohydrate intake of each participant, shown in Figures~\ref{fig:insulin_boxplot}~\&~\ref{fig:carb_boxplot} respectively. This includes all the days between the first and last occurrence of the respective features, which excludes cases such as most of P07's and all of P17's data, where no insulin data is included due to a data export issue. The daily insulin dose varies as a result of numerous factors, including insulin sensitivity, daily activity, and carbohydrate intake. The variation seen in daily insulin doses reflects the different requirements and lifestyles of the participants. The daily carbohydrate shows similar variation across participants, which would fit as carbohydrate intake is linked to higher insulin requirements. P12 has the highest median and maximum daily insulin dose and carbohydrate intake, exemplifying this trend. 

\begin{figure}[tb!]
    \centering
    \includegraphics{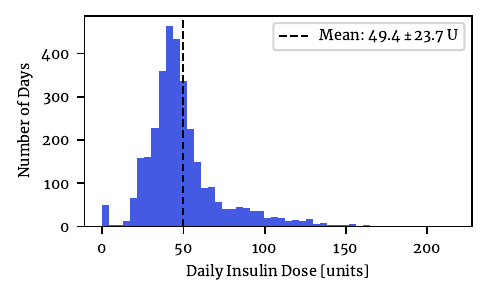}
    \caption{The daily insulin doses across all participants. The majority of days are between 30 and 60 units but there are cases of much higher daily doses and a number of days lacking insulin data.}
    \label{fig:insulin_hist}
\end{figure}

\begin{figure}[bt!]
    \centering
    \includegraphics{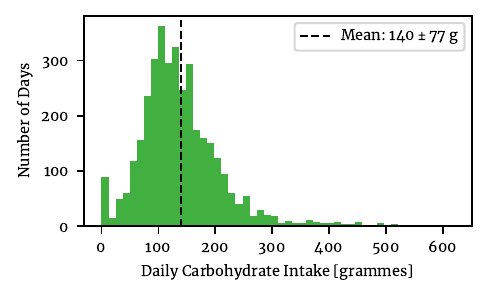}
    \caption{The daily carbohydrate intake across all participants. There are some days lacking data, but the majority show an intake of 80 to 200 grammes. However, due to the collection of carbohydrate data this is likely to be an underestimation.}
    \label{fig:carbs_hist}
\end{figure}

The carbohydrate data is likely to be less reliable as it is recorded when participants enter the value as part of the bolus calculation. This misses cases when the participants eat but do not want to give insulin, for example, to treat hypoglycemia or to counter a drop from activity. The daily insulin dose and carbohydrate intake across all participants are shown in Figures~\ref{fig:insulin_hist} and \ref{fig:carbs_hist}, which also show a similar pattern. There are 48 days of no insulin dose and 84 days of no carbohydrate intake, representing gaps in the participant's data or missing entries. However, this only represents 1.5\% of days within the window that any insulin values were included and 2.3\% of the days within the equivalent window for carbohydrate readings. 

\begin{figure*}[tb!]
    \centering
    \includegraphics{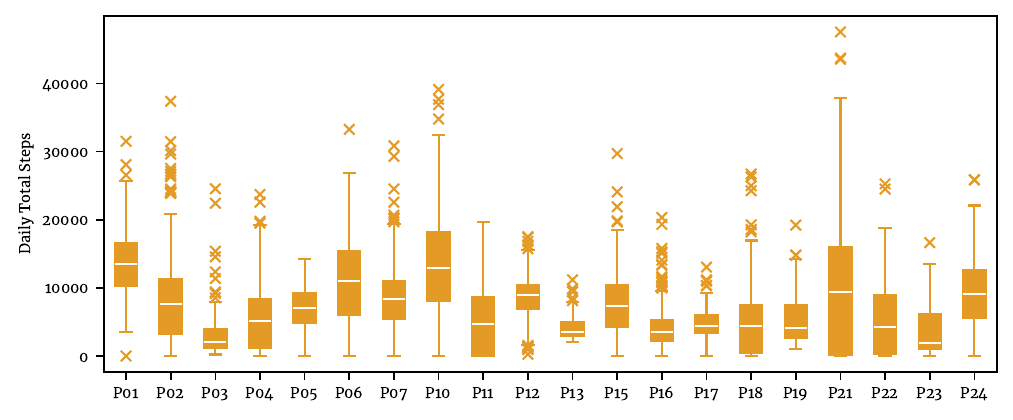}
    \caption{Boxplot of daily step counts for each participant. The activity levels across participants vary, but all have cases of a range of daily step counts against which to compare blood glucose change.}
    \label{fig:steps_boxplot}
\end{figure*}

\begin{figure*}[tb!]
    \centering
    \includegraphics{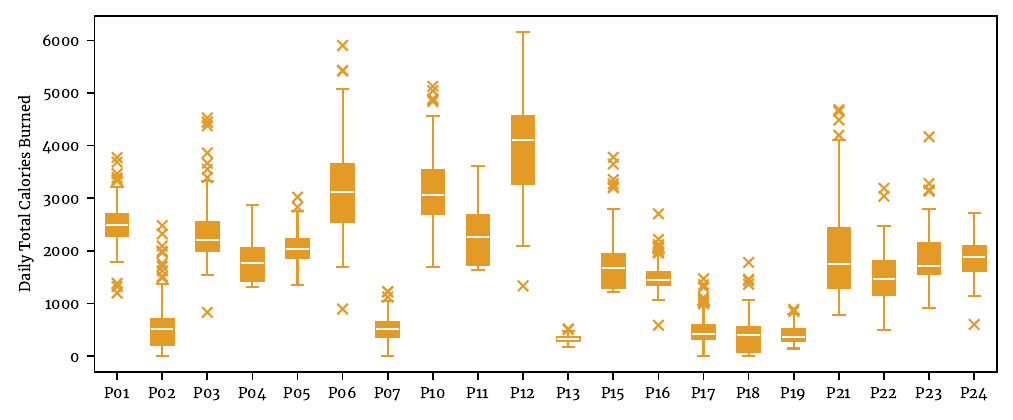}
    \caption{Boxplot of daily calories burned as calculated by the smartwatch. These vary considerably across participants, although the calculations used by Apple Watches and Fitbits are different, which accounts for part of this variation. }
    \label{fig:cals_boxplot}
\end{figure*}

\begin{figure}[tb!]
    \centering
    \includegraphics{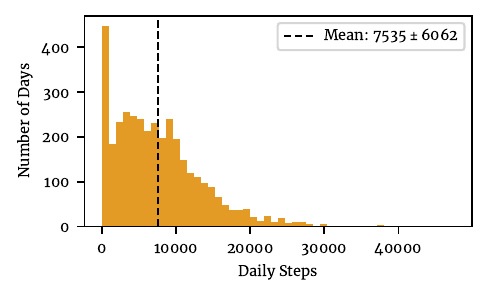}
    \caption{The daily step counts across all participants. The first bin is skewed by 288 days of no steps, which are likely due to the smartwatch not being worn on that day.}
    \label{fig:steps_hist}
\end{figure}

The dataset features heart rate, distance travelled, steps, calories burned, and self-labelled activities from the smartwatch. Steps provide an indication of the base level of activity participants perform, with Figures~\ref{fig:steps_boxplot} and \ref{fig:steps_hist} highlighting the total daily steps for the participants. All participants have a range of daily steps as would be expected and these ranges vary between participants. Of the 3614 days of smartwatch data, there are 288 that have no steps, most likely due to the smartwatch not being worn. This suggests a high level of engagement with the smartwatch. Figure~\ref{fig:cals_boxplot} shows the daily calories burned by each participant as calculated by the smartwatch. P02, P07, P13, P17, P18, and P19 have lower daily calories burned values, which are the 6 participants who used an Apple Watch, compared to a Fitbit for the other participants. The Fitbit assigns a background rate of calories burned, which is recorded even if the watch is not worn, skewing the values. 

\begin{figure}[tb!]
    \centering
    \includegraphics{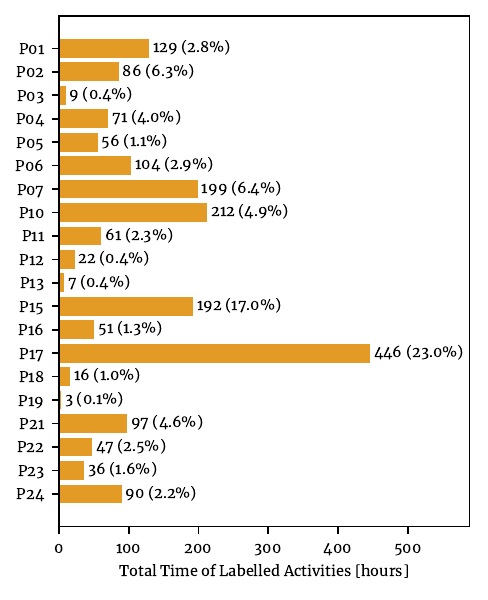}
    \caption{Total number of hours of labelled activity data for each participant. The number of hours for each participant is stated with the percentage of the total number of hours the smartwatch was worn in brackets. There was a range of engagement of this feature.}
    \label{fig:act_bar}
\end{figure}

\begin{figure}[tb!]
    \centering
    \includegraphics{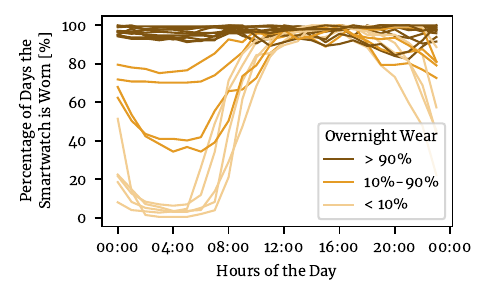}
    \caption{Mean daily wearing patterns of the smartwatch. Each line represents a participant, with the value normalised over the total days of smartwatch data for each participant to account for variation in this total. The participants fell into three pattern of wear: almost always worn overnight (>90\% --- P01, P03, P05, P06, P07, P10, P12, P13, P21, P23, and P24),  sometimes worn overnight (10\%--90\% --- P04, P11, P16, and P22), and almost never worn overnight (<10\% --- P02, P15, P17, P18, and P19).}
    \label{fig:smartwatch_wear}
\end{figure}

The labelled activity feature had mixed engagement across participants, as depicted in Figure~\ref{fig:act_bar}. The variation is due to the manual nature of this data collection, which accurately highlights activity events but is unreliably used by some participants. Examples of the categories that activity was categorised into include `Walk', `Run', `Swim', `HIIT', and `Weights'. The daily smartwatch usage patterns of the participants are highlighted in Figure~\ref{fig:smartwatch_wear}, where heart rate has been used as an indicator for if the smartwatch is being worn. The usage appears to fall into three groups, participants who wore the smartwatch overnight (1) almost always, (2) sometimes, and (3) almost never.

\subsection{Limitations}
Data collected in this study is considered `in the wild', where participants were given no instructions over their use of the smartwatch. Although this is an important feature of the dataset, and every effort has been made to maximise the accuracy of the data, limitations exist due to the `messy' nature of wearable-based health monitoring. The carbohydrate readings are recorded when a participant uses the bolus calculator on their insulin pump to calculate the required dose. This means cases when the participant does not use the bolus calculator or the calculator computes that no dose is required are missed. As a result, there are likely cases where carbohydrate is consumed that impact blood glucose levels, for example, to correct a hypoglycaemic event, that are missed. However, whether this is a limitation is application dependent as while this results in missing data, this is also reflective of real-world use. As a result, it provides a more robust test for prediction and closed-loop algorithms. Due to data being recorded and stored by a range of devices, some of which rely solely on internal clocks, time misalignments in the data are possible. Where possible, this has been corrected for, but there are likely other cases.

The study focuses on young adults as they create a robust test for real-world usage due to the life changes that happen during this period. However, findings from this population may not reflect other age groups as well. Additionally, there are biases in the participant pool involved in the study, as highlighted in Table~\ref{tab:demographics}. There is a high proportion of female and white British participants, so there may be trends that are disproportionately represented in the data and others that are missed that exist in the wider population. The BrisT1D Dataset can be used to provide some insights, but further testing and exploration will be needed to confirm findings across other demographic groups. This dataset is one of a growing pool of publicly available T1D datasets and should be used alongside others where possible.

\section{Re-use Potential}


The BrisT1D Dataset can be utilised for multiple areas of T1D research, including blood glucose prediction \cite{ensemble_bg_predict, data_driven_bg_predict, benchmark_ml_bg_predict, personalised_bg_predict_review, deep_bg_predict}, hypoglycaemia prediction \cite{ml_hypo_predict_review, ml_realtime_hypo_predict}, and physiological data use in closed-loop algorithm development \cite{jacobs_activity_in_closed-loop, jacobs_adaptive_closed-loop_with_activity, turksoy_wearable_data_in_closed-loop}, utilising the quantitative elements of the dataset. The processed state of the device data offers an easy opportunity to analyse each of these areas, and with the regular nature of the data streams, it is easy to generate many scenarios against which models can be trained and tested. The use case for blood glucose prediction is clear, as windows of data can be grouped and then used as inputs to a model that learns to predict the blood glucose levels a set period into the future. It was used for this purpose in the BrisT1D Blood Glucose Prediction Competition \cite{kaggle_brist1d}. Here, the BrisT1D Dataset was reformatted and used to challenge participants to predict blood glucose an hour in the future using the past six hours of device data. The work forms a benchmark performance using the dataset, against which future research can be compared.

The extensive period of data available for many of the participants allows for exploration into methods of personalising to an individual based on their historic T1D management. By training models to capture the individual blood glucose trends of different participants, analysis can be performed on the improvement this makes. Beyond the frequently available features from the smartwatch, which are presented in the processed state device data, there are other features that can be extracted from the restricted side of the dataset containing the raw state device data. Potential avenues for investigation include the impact of sleep, temperature, oxygen saturation, and menstrual cycle on T1D management on a more focused basis. Considering these additional features is a valuable step in exploring and explaining the unexpected patterns experienced by those who manage T1D \cite{unexpected_patterns}.

The dataset can also be used to explore user opinions of using smartwatches as part of T1D management and other user perceptions of T1D management technology using the qualitative components of the dataset \cite{t1d_life_transitions, hcl_interviews, bg_predict_interviews}. A thematic analysis of the interview and focus group transcripts has been completed, exploring the integration of technology into existing self-management ecosystems \cite{ecosystems}. It highlights the potential of smartwatches in T1D management as an information output source, a self-management ecosystem interface, and a data source for other T1D management devices. There is also the potential for mixed-method analysis utilising both sides of the dataset to build more rounded conclusions about smartwatch use in T1D \cite{RN555}. In mixed-methods analysis, comparisons can be made between the participants' comments and their usage of the smartwatch, their activity levels, and their T1D management.

\section{Availability of Source Code and Requirements}


The code used to convert the raw state device data to its processed state is available through GitHub. This code has been released to aid researchers who want to explore the raw state of the device data and the additional information captured during the study. This includes more detailed records of the insulin dosage and the carbohydrate ratios used by participants, and additional smartwatch data features that appear less regularly, such as sleep, device temperature, and breathing variability. The code also allows other researchers to more readily adapt the processing performed to better suit their research needs. Further details about the code and accessing it are:
\begin{itemize}
    \item Project name: BrisT1D Dataset Processing
    \item Project home page: \url{https://github.com/SamAJames/brist1d_processing}
    \item Operating system(s): Platform independent
    \item Programming language: Python (Jupyter Notebooks)
    \item Other requirements: None
    \item License: CC-BY 4.0 \cite{ccby4.0}
\end{itemize}

\section{Data Availability}

The BrisT1D Dataset is split into two parts,
\begin{enumerate}
    \item BrisT1D-Open Dataset \cite{brist1d_open}, and
    \item BrisT1D-Restricted Dataset \cite{brist1d_restricted},
\end{enumerate}
both of which are published on `data.bris', the University of Bristol Data Repository \cite{databris}. The open-access part is readily available to download and licensed under the Creative Commons Attribution 4.0. The restricted-access part of the dataset requires application to access and has a number of access requirements, including institutional affiliation, summary of usage, evidence of ethical approval, and evidence of funding. A data access agreement is then signed between the affiliated institution and the University of Bristol and access to the restricted dataset is granted. These precautions protect the more sensitive raw data that brings with it higher potential for participant identification.

\section{List of Abbreviations}

CGM: Continuous Glucose Monitor; ML: Machine Learning; T1D: Type 1 Diabetes

\section{Consent for Publication}


Ethical approval for the study was received from the University of Bristol Engineering Faculty Research Ethics Committee (Ref: 13065). Within the datasets are versions of the participant information sheet and consent form that were converted to an online form and used in the study.

\section{Competing Interests}

The authors declare that they have no competing interests.

\section{Funding}


This work was supported by the Engineering and Physical Sciences Research Council Digital Health and Care Centre for Doctoral Training at the University of Bristol (UKRI Grant No. EP/S023704/1), seed corn funding from Jean Golding Institute for data science and data-intensive research at the University of Bristol, and funding for the AI for Collective Intelligence Research Hub from the UKRI AI Programme and EPSRC (Grant No. EP/Y028392/1). The National Institute for Health and Care Research Bristol Biomedical Research Centre also funds one of the study’s co-authors (MEGA). The views expressed are those of the authors and not necessarily those of the NIHR or the Department of Health and Social Care.

\section{Author's Contributions}


SGJ: Conceptualization, Data curation, Formal analysis, Funding acquisition, Investigation, Methodology, Project administration, Software, Validation, Visualization, Writing – original draft, and Writing – review \& editing.
MEGA: Conceptualization, Funding acquisition, and Supervision.
AAO: Conceptualization, Funding acquisition, and Supervision.
HE: Conceptualization, Supervision, and Writing – review \& editing.
ZSA: Conceptualization, Funding acquisition, Supervision, and Writing – review \& editing. 
All authors read and approved the final manuscript.

\section{Acknowledgements}






Huge thanks go to the participants for their time and engagement, Breakthrough T1D for their help in participant recruitment and Will for his valuable insights.

\bibliography{paper-refs}

\end{document}